\documentclass[pra,a4paper,showpacs,twocolumn,superscriptaddress,longbibliography]{revtex4-1}

\usepackage{amssymb}
\usepackage{amsmath}
\usepackage{amsfonts}
\usepackage{graphicx}
\usepackage{bm}
\usepackage{color}
\usepackage{multirow}
\usepackage{natbib}
\usepackage{hyperref}
\usepackage[normalem]{ulem}
\usepackage{relsize}

\DeclareMathOperator{\Tr}{Tr}

\DeclareMathOperator{\sgn}{sgn}

\begin{document}

\title{Indirect exchange interaction between magnetic impurities near the helical edge}

\author{V. D. Kurilovich}

\affiliation{Moscow Institute of Physics and Technology, 141700 Moscow, Russia}

\affiliation{Skolkovo Institute of Science and Technology, 143026 Moscow, Russia}

\author{P. D. Kurilovich}

\affiliation{Moscow Institute of Physics and Technology, 141700 Moscow, Russia}

\affiliation{Skolkovo Institute of Science and Technology, 143026 Moscow, Russia}

\author{I. S.~Burmistrov}

\affiliation{L.D. Landau Institute for Theoretical Physics, Kosygina
  street 2, 117940 Moscow, Russia}
  
\affiliation{Moscow Institute of Physics and Technology, 141700 Moscow, Russia}

\begin{abstract}
The indirect exchange interaction between magnetic impurities located in the bulk of a two-dimensional topological insulator decays exponentially with the distance. The indirect exchange interaction for magnetic impurities mediated by the helical states at the edge of the topological insulator demonstrates behaviour which is typical for the Ruderman-Kittel-Kasuya-Yosida interaction in a one-dimensional metal. We have shown that interference between the bulk and edge states in the two-dimensional topological insulator results in existence of the unusual contribution to the indirect exchange interaction which, on the one hand, decays exponentially with a distance at the length scale controlled by the Fermi energy of the edge states and, on the other hand, oscillates with distance along the helical edge with the period determined by the Fermi wave length. We found that this interference contribution to the indirect exchange interaction becomes dominant for such configurations of two magnetic impurities when one of them is situated close to the helical edge whereas the other one is located far away in the bulk.
\end{abstract}


\maketitle

\section{Introduction}
\label{s1}

Two-dimensional (2D) topological insulators have attracted great attention recently due to existence of two spin-momentum locked edge states caused by a strong spin-orbit coupling \cite{Qi-Zhang,Hasan-Kane}. Because of this peculiar structure of the edge states in a topological insulator (TI), a spin current can propagate along the edges. This current is the basis of the quantum spin Hall effect which was predicted theoretically \cite{Kane-Mele, BHZ} and observed experimentally \cite{Konig2007} in HgTe/CdTe quantum wells. One of remarkable features of the helical edge is the perfect transport along it which cannot be suppressed by any perturbation  preserving  the time-reversal symmetry (in the absence of interactions), e.g. by non-magnetic impurities. In the presence of interactions the backscattering is possible which leads to suppression of the edge conductance at finite temperatures \cite{Goldstein2013,Goldstein2014,Gornyi2014}. Moreover, the electron-electron interaction can lead to the edge reconstruction and spontaneous breakdown of the time-reversal protection of the perfect edge transport \cite{Meir2016}.

A local  perturbation which breaks the time-reversal symmetry such as classical magnetic impurities can also provide a source for a spin-flipping scattering of the edge states and, consequently, can affect the transport properties \cite{Maciejko2009,Tanaka2011}. Thus the transport along the helical edge is sensitive to the properties of a system of magnetic impurities distributed not far from the boundary of 2D TI \cite{Maciejko2012,Cheianov2013,Yudson2013,Yudson2015}. For rare magnetic impurities the main source of interaction between them is the indirect exchange interaction (IEI). If magnetic impurities are situated exactly at the edge of a 2D TI the IEI mediated by the helical states has been computed recently  \cite{LeeLee2015}. Its dependence on a distance resembles the Ruderman-Kittel-Kasuya-Yosida (RKKY) interaction for a one-dimensional metal [\onlinecite{Ruderman-Kittel,Kasuya,Yosida}]. We remind that the main features of behaviour of the RKKY interaction with the distance between the impurities are power-law decay and oscillations with period $\pi/k_F$ where $k_F$ denotes the Fermi wave vector of the helical states. The latter favours the formation of a spin glass state at low temperatures. However, the spin structure of the IEI reflects a strong spin-orbit coupling which exists in a 2D TI: there is interaction between the in-plane components of the impurity spins only. We note that considerations of Ref. \cite{LeeLee2015} ignores the fact that the edge states are composed from the electron-like states with the spin 1/2 and the hole-like states with the spin 3/2 as well as the presence of bulk states.  

In the opposite limit, when the magnetic impurities are located deep in the bulk of a 2D TI, typical semiconductor behaviour of the IEI can be expected. The IEI in three-dimensional (3D) semiconductors with the chemical potential pinned to the gap was first studied by Bloembergen and Rowland [\onlinecite{Bloembergen1955}]. At low temperatures the IEI between magnetic impurities was found to decay exponentially with the distance. In the simplest case of isotropic spectrum with minimum (maximum) of the conduction (valence) band at the $\Gamma$ point the sign of the IEI is constant and ferromagnetic ordering of the magnetic impurities is favoured (see Ref. [\onlinecite{Korenblit1978},\onlinecite{Abrikosov1980}] for a review). 

Recently, the IEI between magnetic impurities situated far away from the edges of the 2D TI based on CdTe/HgTe/CdTe quantum well (QW) has attracted a theoretical interest \cite{Kernreiter2016,Kurilovichi2016}. In this case the IEI has rather complicated  spin structure and decays exponentially with the distance at low temperatures provided that the chemical potential is pinned to the gap. It involves anisotropic XXZ Heisenberg interaction, magnetic pseudodipole interaction, and Dzyaloshinsky-Moriya interaction \cite{Kurilovichi2016}. Such spin structure is typical for systems with a strong spin-orbit coupling, e.g. for magnetic impurities at the surface of a 3D TI \cite{Zhang2009I,Ye2010,Garate-Franz,Biswas-Balatsky,Abanin-Pesin,Zhu2011,Efimkin-Galitski,Litvinov2016}. The presence of inversion asymmetry  of the CdTe/HgTe/CdTe quantum well \cite{Dai2008,Winkler2012,Weithofer-Recher,Tarasenko2015} results in even more complicated spin structure of the IEI, which becomes non-invariant under rotations in the plane of the QW. Besides, oscillations of the IEI with the distance appear \cite{Kurilovichi2016}.

In this paper we study theoretically the indirect exchange interaction between magnetic impurities situated near the helical edge of a 2D topological insulator, based on the CdTe/HgTe/CdTe QW. We concentrate on the case of low temperatures and the chemical potential lying within the energy gap of the bulk spectrum. Contrary to all previous studies we take into account simultaneously the edge and bulk states; the latter are modified by the presence of the edge.  We find the following  interesting features of the indirect exchange interaction in a 2D TI:
\begin{itemize}
\item[(i)] The IEI between magnetic impurities can be split up into three parts: contribution of the Bloembergen-Rowland type due to the bulk states which decays exponentially with the distance; contribution due to the edge states which resembles RKKY interaction in a one-dimensional metal; contribution due to an interference between bulk and edge states. Depending on positions of the magnetic impurities the IEI is dominated by one among three contributions. 

\item[(ii)]  The edge state contribution to the IEI involves in-plane spin components only, in agreement with Ref. \cite{LeeLee2015}. 
\item[(iii)] The interference term in the IEI decays exponentially with the distance, but the decaying length depends explicitly on the position of the chemical potential within the bulk gap. 
\end{itemize} 

The outline of the paper is as follows.  In Sec. \ref{Sec:BHZ} we remind a reader the Bernevig-Huges-Zhang Hamiltonian for 2D electron and hole states in the (001) symmetric CdTe/HgTe/CdTe QW and formulate the problem. In Sec. \ref{Sec:MGF} we study the structure of bulk and edge states and compute the Matsubara Green's function. The results for the IEI are presented in Sec. \ref{Sec:IEI}. The discussion of the obtained results and conclusions are given in Sec. \ref{Sec:DisConc}. The technical details of derivation of different contributions to the IEI interaction are presented in Appendices.

\section{The model \label{Sec:BHZ}}

We start from Bernevig-Hughes-Zhang Hamiltonian which can be used to describe low-energy physics of electron and hole states in a 2D TI based on the (001) CdTe/HgTe/CdTe QW \cite{BHZ}. Written in the basis of spatially quantized states of the QW which are commonly denoted as $\vert E_1,+\rangle$, $\vert H_1,+\rangle$, $\vert E_1,-\rangle$, $\vert H_1,-\rangle$ (for details on structure of these states see \cite{BHZ,Rothe,Tarasenko2015}), it has the following form:
\begin{equation} 
H_{\rm BHZ}= \varepsilon(k) +
\begin{pmatrix} 
M(k)&Ak_{+}&0&\Delta\\ 
Ak_{-}&-M(k)&-\Delta&0\\ 
0&-\Delta&M(k)&-Ak_{-}\\ 
\Delta&0&-Ak_{+}&-M(k)\\ 
\end{pmatrix}  .
\label{Eq: BHZ}
\end{equation}
Here we introduce
\begin{equation}
\varepsilon(k)=C-D(k_x^2+k_y^2), \quad 
M(k)=M-B(k_x^2+k_y^2) .
\end{equation} 
The parameters $A$, $B$, $C$, $D$, $\Delta$ and $M$ depend on the width $d$ of the QW. The term $\Delta$ describes the interface and bulk inversion asymmetry and, generally, can be comparable to the gap $M$ \cite{Tarasenko2015}. 

As it was shown in Ref. \cite{Kurilovichi2016}, the terms quadratic in the momentum in the Hamiltonian \eqref{Eq: BHZ} are not important for the calculation of the IEI. Therefore we shall consider a simplified model given by the Hamiltonian \eqref{Eq: BHZ} in which we set $B=D=0$. The inversion asymmetry term $\Delta$ results in oscillating dependence of the IEI on the distance between magnetic impurities situated in the bulk of the QW \cite{Kurilovichi2016}. In order to simplify the calculations of the IEI in the presence of the helical  edge we neglect $\Delta$ in the present paper.  Thus the Hamiltonian we shall work with is given by the following expression:
\begin{equation}\label{Eq: BHZ-lin}
H=
\begin{pmatrix} 
M&Ak_{+}&0&0\\ 
Ak_{-}&-M&0&0\\ 
0&0&M&-Ak_{-}\\ 
0&0&-Ak_{+}&-M\\ 
\end{pmatrix} .
\end{equation} 

The Hamiltonian of a magnetic impurity with the spin $S$ situated at some point $\{x_0,y_0,z_0\}$ within the (001) QW reads \cite{Kurilovichi2016}: 
\begin{equation}
\mathcal{V}_{\rm imp} = \mathcal{J} \delta(x-x_0) \delta(y-y_0) ,
\end{equation}
where the matrix 
\begin{equation}
\mathcal{J} =
\begin{pmatrix}
J_1 S_z & -iJ_0 S_+ & J_{m} S_{-} &0\\
iJ_0S_{-} & J_2 S_z &0 &0\\
J_{m} S_{+} &0 & -J_1 S_z & - iJ_0 S_{-}\\
0 & 0 &  iJ_0 S_+ & -J_2 S_z   
\end{pmatrix} 
\label{eq:Jmat}
\end{equation}
describes interaction with electron and hole states $\vert E_1,+\rangle$, $\vert H_1,+\rangle$, $\vert E_1,-\rangle$, $\vert H_1,-\rangle$. The coupling constants $J_0$, $J_1$, $J_2$ and $J_m$ depend on $z_0$ and are determined by  the envelope functions of spatially quantized states in the QW (see Ref. \cite{Kurilovichi2016} for the details).

\section{The Matsubara Green's function \label{Sec:MGF}}

\begin{figure}[t]
\centerline{\includegraphics[width=0.43\textwidth]{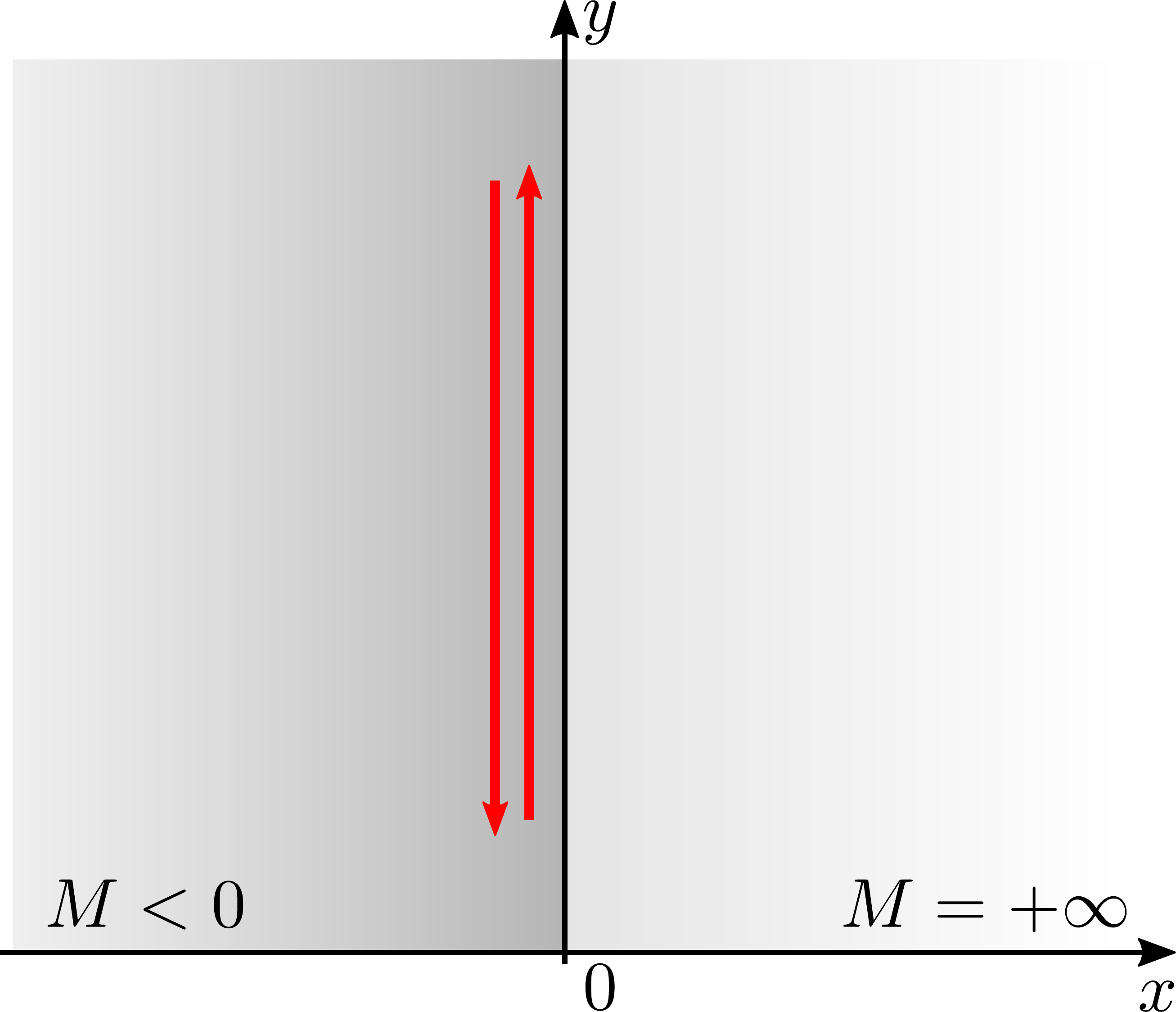}}
 \caption{(Color online) Topological insulator (shaded gray) resides at $x<0$. In this region the gap $M$ is finite and negative. At $x>0$ the gap is positive and infinite. A pair of helical states (red lines) propagate along the edge.}
 \label{H-edge}
\end{figure}

In order to evaluate the expression for the IEI it is convenient to use the Green's function approach. Thus, we start from examing the Green's function for a 2D TI with a straight boundary situated at $x=0$. We adopt the approach of Ref. \cite{VolkovPankratov} and assume that the gap $M$ is a function of $x$ such that $M(x)$ equals a negative constant for $x<0$ and $M(x)=+\infty$ for $x>0$, see Fig.~\ref{H-edge}. 

As we consider a system of non-interacting electrons described by the Hamiltonian \eqref{Eq: BHZ-lin} in the presence of the boundary at $x=0$, it is necessary to take into account several important features. At first, there exist the edge states localized near the boundary which contribute to the Green's function. Secondly, the structure of the bulk states in the presence of the boundary differs from the case of an infinite sample in which $k_x$ is a good quantum number. It is convenient to evaluate the expression for the Green's function using the Lehmann's representation:
\begin{equation}\label{Eq:Lehmann}
\mathcal{G}(i\varepsilon_n, \bm{r}, \bm{r}^\prime)=\sum_m \frac{\psi_{m}(\bm{r}) \psi^\dagger_{m}(\bm{r}^\prime)}{i\varepsilon_n+\mu-\epsilon_m} ,
\end{equation}
where $m$ enumerates eigenstates $\psi_m(\bm{r})$ of the Hamiltonian \eqref{Eq: BHZ-lin} with an energy $\epsilon_m$. The  chemical potential is denoted by $\mu$ and the Matsubara fermionic energy $\varepsilon_n =\pi T (2n+1)$.

The Lehmann's representation suggests to split the Green's function into two parts: $\mathcal{G}=\mathcal{G}_{\mathrm{edge}}+\mathcal{G}_{\mathrm{bulk}}$. In $\mathcal{G}_{\mathrm{edge}}$ ($\mathcal{G}_{\mathrm{bulk}}$) summation over the edge (bulk) states is performed only.

\subsection{The Green's function of the edge states}

There exists a pair of the edge states connected via the time-reversal symmetry. For a given $k_y$ one state is associated with the upper block of $4\times 4$ Hamiltonian \eqref{Eq: BHZ-lin} and the other -- with the lower one. They have the following form
\begin{gather}
\psi_{\mathrm{edge},\uparrow}(k_y, \bm{r})=\begin{pmatrix}
1\\i\\0\\0\end{pmatrix}\frac{e^{ik_y y}}{\sqrt{2\pi \xi}}e^{x/\xi} \theta(-x), \\
\psi_{\mathrm{edge},\downarrow}(k_y, \bm{r})=\begin{pmatrix}
0\\0\\1\\-i\end{pmatrix}\frac{e^{ik_y y}}{\sqrt{2\pi \xi}}e^{x/\xi}  \theta(-x),
\end{gather}
where  $\xi = A/|M|$ and $\theta(z)$ denotes the Heaviside step function. The energy spectrum of the edge states is linear in the momentum: $\epsilon_{\mathrm{edge},\uparrow / \downarrow}(k_y)=\pm A k_y$.
Integrating over the momentum $k_y$, we find
\begin{equation}
\mathcal{G}_{\mathrm{edge}}=G^{\uparrow}_{\mathrm{edge}}+G^{\downarrow}_{\mathrm{edge}} ,
\label{eq:G:edge:1}
\end{equation}
where
\begin{gather}
G^{\uparrow/\downarrow}_{\mathrm{edge}}(i\varepsilon, \bm{r}, \bm{r}^\prime)=\pm \frac{i|M|}{A^2} e^{(x+x')/\xi\pm (\varepsilon-i\mu) (y-y')/A} \theta(-x)\notag\\\times \theta(-x^\prime)\bigl [ \theta (y-y') \theta (\mp\varepsilon) - \theta (y'-y) \theta (\pm\varepsilon)\bigr ]\Gamma_{\pm} .
\label{eq:G:edge:2}
\end{gather}
Here the matrices $\Gamma_\pm$ are defined as follows
\begin{equation}
\Gamma_+=\begin{pmatrix}
1 & -i & 0 & 0 \\
i & 1 & 0 & 0 \\
0& 0& 0& 0 \\
0& 0& 0& 0
\end{pmatrix}, \qquad 
\Gamma_-=\begin{pmatrix}
0& 0& 0& 0 \\
0& 0& 0& 0 \\
0 & 0 & 1 & i  \\
0 & 0 & -i & 1  \\
\end{pmatrix} .
\end{equation}

\subsection{The Green's function of the bulk states}

Now we discuss the structure of the bulk states in the presence of the boundary as well as the bulk part of the Green's function. There are four bulk states for the Hamiltonian \eqref{Eq: BHZ-lin}. Two of them have positive energy: $\epsilon^+_{\mathrm{bulk},\uparrow/\downarrow}(k)={\cal{E}}(k)$, where ${\cal{E}}(k)=\sqrt{M^2+A^2 k^2}$, and two have negative energy: $\epsilon^-_{\mathrm{bulk},\uparrow/\downarrow}(k)=-{\cal{E}}(k)$. It is convenient to introduce the following functions
\begin{gather}
f^{\pm}_x(\bm{k}) = \frac{\left( A k_\pm \pm i\left( {\cal{E}}(k)\mp|M|\right)\right) e^{ik_x x} + \mathrm{c.c.}}{2\sqrt{{\cal{E}}(k)({\cal{E}}(k)+Ak_y)}} ,
\end{gather}
where $k_\pm = k_x \pm i k_y$.
In terms of these functions one can present the bulk eigenstates as
\begin{gather}
\psi_{\mathrm{bulk}, \uparrow}^{\pm}(\bm{r})=
\begin{pmatrix}
\pm f^\pm_x(\pm\bm{k})\\
\pm i f^\mp_x(\pm\bm{k}) \\
0\\
0
\end{pmatrix}\frac{e^{ik_y y}}{2\pi},\notag \\
\psi_{\mathrm{bulk}, \downarrow}^\pm(\bm{r})=
\begin{pmatrix}
0\\
0\\
\mp f^\pm_x(\mp\bm{k})\\
\pm if^\mp_x(\mp\bm{k})
\end{pmatrix}\frac{e^{ik_y y}}{2\pi} .
\label{eq:bb:wf}
\end{gather}
The upper index `$\pm$' indicates whether electron (+) or hole (-) band is concerned.
The Green's function of the bulk states can be written in the following form:
\begin{gather}
\mathcal{G}_{\mathrm{bulk}}(i\varepsilon, \bm R, \bm R')
= \sum_{s=\pm}  \int \frac{d^2\bm{k}}{(2\pi)^2}  e^{ik_y (y - y')} \frac{\theta(k_x){\cal{B}}_s(\bm k, x, x')}{i\varepsilon+\mu-s{\cal{E}}(k)} ,
\label{eq: bulkgf}
\end{gather}
where ${\cal{B}}_{s}$ is the following $4\times 4$ block diagonal matrix:
\begin{align}
{\cal{B}}_{s}(\bm{k},x,x') & = 
\begin{pmatrix}
\hat{b}_s(s\bm{k},x,x') & \hat{0}\\
\hat{0} & \hat{b}_{s}^T(-s \bm{k},x',x)
\end{pmatrix}
,\notag \\
\hat{b}_s(\bm{k},x,x') & =
\begin{pmatrix}
f_x^s(\bm k)f_{x'}^s(\bm k)&-if_x^s(\bm k)f_{x'}^{-s}(\bm k)\\
if_x^{-s}(\bm k)f_{x'}^s(\bm k)&f_x^{-s}(\bm k)f_{x'}^{-s}(\bm k)
\end{pmatrix} .
\end{align}
The superscript `T' denotes the matrix transposition.

\section{The indirect exchange interaction\label{Sec:IEI}}

Let us now turn to the calculation of the indirect exchange interaction.
To the second order in $\mathcal{J}$ the IEI is given by a polarization operator diagram. The corresponding effective Hamiltonian that describes the interaction of two magnetic impurities situated at points $\bm{r}_A=\{\bm{R}_A,z_A\}$ and $\bm{r}_B=\{\bm{R}_B,z_B\}$ can be written as
\begin{equation}
{H}_{\rm IEI} = T\sum_{\varepsilon_n} \Tr  \mathcal{J}^A \mathcal{G}(i\varepsilon_n, \bm{R}_A, \bm{R}_B) 
 \mathcal{J}^B \mathcal{G}(i\varepsilon_n, \bm{R}_B, \bm{R}_A)  .
 \label{H-IEI}
\end{equation}
In this paper  we focus on the case of the zero temperature only. Thus, we replace the summation over Matsubara frequencies by the integral over the energy. The superscript $A$ ($B$) in $\mathcal{J}^A$ ($\mathcal{J}^B$) indicates that the matrix \eqref{eq:Jmat} is evaluated at the position $z_A$ ($z_B$). 

Since the Hamiltonian (\ref{H-IEI})  involves a product of two Green's functions under the sign of $\Tr$ and each of them is a sum of the edge and bulk contributions, one can decompose the IEI as a sum of the following three terms:
\begin{gather}
H_{\mathrm{IEI}}=H_{\mathrm{IEI}}^{\mathrm{bulk}}+H_{\mathrm{IEI}}^{\mathrm{edge}}+H_{\mathrm{IEI}}^{\mathrm{int}}.
\end{gather}
The first term in the right hand side of this equation, $H_{\mathrm{IEI}}^{\mathrm{bulk}}$, is related to the bulk states only: it involves the product of two bulk Green's function $\mathcal{G}_{\mathrm{bulk}}$.  The second term, $H_{\mathrm{IEI}}^{\mathrm{edge}}$, is related to the edge states. It contains the edge Green's function, $\mathcal{G}_{\mathrm{edge}}$, only. The last term, $H_{\mathrm{IEI}}^{\mathrm{int}}$, describes the interference between the bulk and the edge states and involves the edge and bulk Green's functions simultaneously.

Before proceeding with the results for the three different contributions to the IEI, let us briefly discuss the notations.
Hereinafter, we denote $\bm{R}_{A/B}\equiv \left(x_{A/B},y_{A/B}\right)$, $\bm{R} = \bm{R}_{A}-\bm{R}_{B}\equiv \left(x_{AB},y_{AB}\right)$, $\bm{n} = \bm{R}/R$, $\overline{x}_{AB} \equiv x_A+x_B$,  $\overline{\bm{R}}= \left(\overline{x}_{AB},y_{AB}\right)$, and $\bm{\nu} = \overline{\bm{R}}/\overline{{R}}$, where  $\overline{R}=\sqrt{y_{AB}^2+\overline{x}_{AB}^2}$. In addition, we assume below  that $y_{AB}>0$.

\subsection{Bulk and edge contributions to the IEI}

The part of the IEI mediated by the bulk states has the complex form with non-trivial spin structure in general. In the absence of the boundary, the asymptotic expression for the IEI at the distances $R\gg \xi$ reads \cite{Kurilovichi2016}:
\begin{align}
{H}^{\rm bulk}_{\rm IEI}  & = \frac{1}{|M| \xi^4} \left( \frac{\xi}{4\pi R} \right)^{3/2}e^{-2R/\xi} \Bigl [  J_m^A J_m^A  \bigl ( \bm{S}_\|^A \cdot \bm{S}_\|^B \bigr ) +
\notag \\
& + 2  \Bigl ( J_0^A J_z^B \bigl ( \bm{S}_\|^A \cdot \bm{n} \bigl ) S_z^B - J_z^A J_0^B
S_z^A \bigl ( \bm{S}_\|^B \cdot \bm{n} \bigl )\Bigr ) -\notag \\
& -
4 J_0^A J_0^B \bigl ( \bm{S}_\|^A \cdot \bm{n} \bigl ) \bigr (\bm{S}_\|^B \cdot \bm{n}\bigr ) + J_z^A J_z^B S_z^A S_z^B   \Bigr ]  ,
\label{eq:bulk-IEI}
\end{align}
where $J^{A/B}_z = J^{A/B}_1 + J^{A/B}_2$. In the presence of the boundary the bulk states acquire non-trivial structure, Eqs. \eqref{eq:bb:wf}, that complicates the form of the IEI. The large distance asymptote of the full expression is presented in Appendix \ref{App1}. Additional terms, which appear, can be interpreted as the interaction between a magnetic impurity and the mirror image of the other impurity with respect to the boundary. These additional terms decay in a different way:  $\sim \exp[-(R+\overline{R})/\xi]$ and $\sim \exp (-2\overline{R}/\xi)$. Therefore, in the presence of the boundary Eq. \eqref{eq:bulk-IEI} is valid provided the following inequalities are satisfied:
\begin{equation}
|x_A|, |x_B| \gg \xi. 
\end{equation}
The result \eqref{eq:bulk-IEI} has been derived for the zero temperature. At finite temperature, this result is valid provided the following inequality holds \cite{Kurilovichi2016}:
\begin{equation}
\frac{M^2}{T^2} \min \left \{1, \frac{T}{|\mu|} \left (1 - \frac{\mu^2}{M^2} \right )\right \} \gg \frac{R}{\xi} \gg 1 .
\end{equation}

The expression for the contribution to the IEI due to the edge states only can be derived exactly at $T=0$ within the help of Eqs. \eqref{eq:G:edge:1} and \eqref{eq:G:edge:2}. The result is as follows
\begin{gather}
H^{\mathrm{IEI}}_{\mathrm{edge}}=-  \frac{e^{2\overline{x}_{AB}/\xi}}{2\pi y_{AB} |M| \xi^3} J_{m}^A J_{m}^B \Bigl [ \cos \left(2 k_F y_{AB}\right) \left( \mathbf{S}^A_{||} \cdot \mathbf{S}^B_{||} \right)+\notag \\
+ \sin \left( 2 k_F y_{AB}\right) \left[\mathbf{S}^A \times \mathbf{S}^B \right]_z \Bigr ] ,
\label{Eq: IEI-edge}
\end{gather}
where $k_F= \mu/A$ denotes the Fermi wave vector of the edge states. We mention that our result \eqref{Eq: IEI-edge} for magnetic impurities situated exactly at the boundary, $x_A=x_B=0$ coincides with the result derived in Ref. \cite{LeeLee2015}. We note that a magnetic impurity situated away from the boundary interacts by means of the helical edge states with the mirror image of the other impurity with respect to the boundary only. This is the consequence of the absence of the translational invariance perpendicular to the boundary (along the $x$ axis). The dependence of the IEI mediated by the edge states has a typical one-dimensional metallic behaviour: it decays inversely proportional to the distance and oscillates in space with the period $\pi/k_F$. A feature of the result \eqref{Eq: IEI-edge} is that the edge contribution to the IEI couples in-plane components of the impurity spins only. We will discuss later why this behaviour may be crucial for the IEI if the on-site spin anisotropy is present. 

At finite temperature the result \eqref{Eq: IEI-edge} is valid for not too large values of $y_{AB}$:
\begin{equation}
|y_{AB}|/\xi \ll |M|/(\pi T) . 
\end{equation}

\subsection{Interference contribution to the IEI \label{SubSec: Int Int}}

The evaluation of the interference contribution to the IEI is complicated for an arbitrary disposition of the magnetic impurities. In order to obtain analytic results we consider two limiting cases: (i) at least one of the impurities is situated far away from the boundary; (ii) both impurities are located near the boundary. In addition, we assume that the chemical potential is not pinned to the center of the bulk gap, $\mu \neq 0$. For $\mu=0$ the interference contribution has the same decaying length as the bulk contribution and, thus, is of no special interest. 

 \subsubsection{A magnetic impurity away from the boundary}

We start from the case when at least one of the impurities is located far away from the boundary, $|x_A| \gg \xi$ or $|x_B| \gg \xi$. In order to obtain the expression for the interference contribution to the IEI in this case, it is convenient to separate the bulk Green's function into two parts:
\begin{equation}
{\cal{G}}_\mathrm{bulk} = {\cal{G}}^{\mathrm{i}}_\mathrm{bulk}+{\cal{G}}^{\mathrm{ni}}_\mathrm{bulk},
\end{equation}
where ${\cal{G}}^{\mathrm{i}}_\mathrm{bulk}$ (${\cal{G}}^{\mathrm{ni}}_\mathrm{bulk}$) is the translationally invariant (noninvariant) part. The translationally invariant part of the bulk Green function depends only on the relative position of the impurities, while the translationally noninvariant part is suppressed when both impurities are far away from the edge.

Splitting the bulk Green's function into translationally invariant and noninvariant parts allows us to express the interference contribution to the IEI as a sum of the two terms: 
\begin{gather}
H^{\mathrm{int}}_{\mathrm{IEI}} = H^{\mathrm{int,\:i}}_{\mathrm{IEI}}+H^{\mathrm{int,\:ni}}_{\mathrm{IEI}},
\label{eq: sum-h}
\end{gather}
where the former is given by the product of the edge Green's function and the invariant part of the bulk Green's function, while the latter can be expressed as the product of the edge Green's function and the noninvariant part of the bulk Green's function.

For the sake of convenience, we introduce a set of coupling constants $K$ defined as follows
\begin{equation}\label{eq: int}
H^{\mathrm{int,\:i/ni}}_{\mathrm{IEI}} = \sum\limits_{a,b=x,y,z}S^A_a K^\mathrm{int,\: i/ni}_{ab} S^B_b .
\end{equation}
The large distance asymptote of the matrix $K^\mathrm{int,\:i}$ which determines the invariant part of the interference contribution to the IEI  is given by the following expressions (see Appendix \ref{App2}):
\begin{widetext}
\begin{equation}
\label{eq: K-i}
\begin{split}
    K_{xx}^\mathrm{int, i} & = 2F_\mu(\bm R) g_\mu(y_{AB})\left[(1-\sin \theta_\mu)J_{m}^AJ_{m}^B-2\left(\sin \theta_\mu-i \cos \theta_\mu n_y\right)J_0^AJ_0^B\right]+\mathrm{c.c.},\\
    K_{yy}^\mathrm{int, i} & = 2F_\mu(\bm R)g_\mu(y_{AB})\left[(1-\sin \theta_\mu)J_{m}^AJ_{m}^B-2\left(\sin \theta_\mu+i \cos \theta_\mu n_y\right)J_0^AJ_0^B\right]+\mathrm{c.c.},\\
    K_{zz}^\mathrm{int, i} & = 2F_\mu(\bm R)g_\mu(y_{AB})\Bigl[(1-\sin \theta_\mu)J_{1}^AJ_{1}^B-(1+\sin \theta_\mu)J_{2}^AJ_{2}^B+\cos \theta_\mu  n_+ J_1^AJ_2^B- \cos \theta_\mu  n_- J_2^AJ_1^B)\Bigr]+\mathrm{c.c.},\\
    K_{xy}^\mathrm{int, i} & =2iF_\mu(\bm R)g_\mu(y_{AB})\Bigl[(1-\sin \theta_\mu)J_m^AJ_m^B+2J_0^AJ_0^B\left(1-\cos \theta_\mu n_x\right)\Bigr]+\mathrm{c.c.},\\
    K_{xz}^\mathrm{int, i} & =2F_\mu(\bm R)g_\mu(y_{AB})\Bigl[ (1-\sin \theta_\mu-\cos \theta_\mu  n_-) J_0^AJ_1^B-(1+\sin \theta_\mu-\cos \theta_\mu  n_+)J_0^AJ_2^B \Bigr ] +\mathrm{c.c.},\\
    K_{yz}^\mathrm{int, i} & =-2iF_\mu(\bm R)g_\mu(y_{AB})\Bigl[  (1-\sin \theta_\mu+\cos \theta_\mu  n_-)J_0^AJ_1^B +(1+\sin \theta_\mu+\cos \theta_\mu  n_+)J_0^AJ_2^B \Bigr]+\mathrm{c.c.} .
    \end{split}
\end{equation}
\end{widetext}
Here $n_\pm = n_x\pm i n_y$ and the phase $\theta_\mu$ satisfies the following relations:
\begin{equation}\label{eq: thetamu}
\sin \theta_\mu = \frac{\mu}{|M|}, \quad \cos \theta_\mu = \sqrt{1-\frac{\mu^2}{M^2}} .
\end{equation}
The dimensionless function $g_\mu(y)$ is defined as follows
\begin{gather}
g_\mu(y) = \frac{e^{ik_F y}}{\sin \theta_\mu + i \cos \theta_\mu n_y} .
\end{gather}
The function 
\begin{gather}
F_\mu(\bm R) =  \frac{\sqrt{\cos{\theta_\mu}}}{2 |M| \xi^4} \left(\frac{\xi}{2\pi R}\right)^{3/2}e^{\overline{x}_{AB}/\xi-R/\xi_\mu} 
\label{eq:Fmu}
\end{gather}
determines the spatial decay of the translationally invariant part of the interference contribution to the IEI. The corresponding decay length scale is given by
\begin{equation}\label{eq: xi}
\xi_\mu=\xi/\sqrt{1-\mu^2/M^2} . 
\end{equation}
The remaining set of matrix elements of $K^\mathrm{int,\: i}_{ab}$ can be obtained from the ones presented above: $K^\mathrm{int,\: i}_{yx}$, $K^\mathrm{int,\: i}_{yz}$ and $K^\mathrm{int,\: i}_{zx}$ can be read from $K^\mathrm{int,\: i}_{xy}$, $K^\mathrm{int,\: i}_{zy}$ and $K^\mathrm{int,\: i}_{xz}$, respectively,  upon change of $\bm{R}$ to $-\bm{R}$ and swap of subscripts $A$ and $B$.

Appearance of the decay length $\xi_\mu$ in the exponent is somewhat unexpected, as it depends explicitly on the position of the chemical potential $\mu$ and, thus, 
can be electrically tuned. Moreover $\xi_\mu$ diverges as the chemical potential approaches the bulk spectrum. In addition to the term  $-R/\xi_\mu$, there is the term $\overline{x}_{AB}/\xi$ in the exponent of $F_\mu(\bm R)$. Besides the features mentioned above, the matrix elements $K^\mathrm{int, i}_{ab}$ oscillate with the distance along the edge with a period $2\pi /k_F$ which is two times longer than the period of oscillations of the contribution to the IEI mediated by the edge states only. These particular features of the interference contribution to the IEI are descendants of the properties of the edge Green's function.

The result \eqref{eq: K-i} is obtained in the saddle point approximation and is valid for the large distances
\begin{equation} \label{eq: ineq}
R \gg \xi_\mu/\sin^2 \theta_\mu .
\end{equation}
Note, that the right hand side of this inequality diverges at $\mu = \pm |M|$, i.e. when the chemical potential touches the bulk bands. At finite temperatures the result \eqref{eq: K-i} is valid for not too large distance between the impurities:
\begin{equation}
(x_{AB}^2 \sin^2 \theta_\mu + y_{AB}^2)^{1/2} / \xi \ll |M|/(\pi T) .
\label{Tcond: K-i}
\end{equation}

The matrix elements of  $K_{ab}^\mathrm{int,\:ni}$ of the noninvariant part of the interference contribution to the IEI are given by the following expressions (see Appendix \ref{App2}):
\begin{widetext}
\begin{equation}\label{eq: K-n-i}
\begin{split}
    K_{xx}^\mathrm{int,\:ni} & = -2 F_\mu(\overline{\bm R})\overline{g_\mu}(y_{AB}) \left[ 
    \Bigl (u_\mu-u_{-\mu}^*+iv_\mu+i v_{-\mu}^*\Bigr ) J_0^AJ_0^B   + u_\mu^* J_m^A J_m^B
    \right ]+\mathrm{c.c.},  \\
    K_{yy}^\mathrm{int,\:ni} & = -2F_\mu(\overline{\bm R})\overline{g_\mu}(y_{AB})\left[ 
    \Bigl (u_\mu-u_{-\mu}^*-iv_\mu-i v_{-\mu}^*\Bigr ) J_0^AJ_0^B   + u_\mu^* J_m^A J_m^B
    \right ]+\mathrm{c.c.},\\
    K_{zz}^\mathrm{int,\:ni} & = -2 F_\mu(\overline{\bm R})  \overline{g_\mu}(y_{AB}) \Bigl[u_\mu J_1^AJ_1^B
  -u^*_{-\mu} J_2^AJ_2^B  + i v_\mu J_1^AJ_2^B+i v_{-\mu}^* J_2^AJ_1^B \Bigr]+\mathrm{c.c.},\\
    K_{xy}^\mathrm{int,\:ni} & =-2i F_\mu(\overline{\bm R})\overline{g_\mu}(y_{AB})\left[ 
    \Bigl (u_\mu+u_{-\mu}^*-iv_\mu+i v_{-\mu}^*\Bigr ) J_0^AJ_0^B   + u_\mu^* J_m^A J_m^B
    \right ]+\mathrm{c.c.},\\
    K_{xz}^\mathrm{int,\:ni} & =-2F_\mu(\overline{\bm R})\overline{g_\mu}(y_{AB})\Bigl[ (u_\mu+i v_{-\mu}^*) J_0^A J_1^B-(u_{-\mu}^*-i v_{\mu}) J_0^AJ_2^B\Bigr]+\mathrm{c.c.},\\
    K_{yz}^\mathrm{int,\:ni} & =2 i F_\mu(\overline{\bm R})\overline{g_\mu}(y_{AB})\Bigl [ (u_\mu-i v_{-\mu}^*) J_0^A J_1^B+(u_{-\mu}^*+i v_{\mu}) J_0^AJ_2^B \Bigr ]+\mathrm{c.c.} 
\end{split}
\end{equation}
\end{widetext}
Here  $\nu_\pm = \nu_x\pm i \nu_y$ and we introduce the dimensionless functions  
\begin{gather}
\overline{g_\mu}(y) = \frac{e^{ik_F y}}{\sin \theta_\mu + i \cos \theta_\mu \nu_y} 
\end{gather}
and
\begin{equation}
\begin{split}
u_\mu & = -\frac{1}{2} \frac{(
\nu_+\cos\theta_\mu+\sin\theta_\mu-1)^2}{\sin \theta_\mu+ i \nu_y \cos \theta_\mu} , \\
v_\mu &= \frac{i\cos \theta_\mu(\cos\theta_\mu-\nu_x-i \nu_y \sin \theta_\mu)}{\sin \theta_\mu+ i \nu_y \cos \theta_\mu} .
\end{split}
\end{equation}
The elements $K_{yx}^\mathrm{int,\:ni}$ and $K_{yz}^\mathrm{int,\:ni}$ are equal to $-K_{xy}^\mathrm{int,\:ni}$ and $-K_{zy}^\mathrm{int,\:ni}$ after interchange of subscripts $A$ and $B$ as well as $v_\mu$ and $v^*_{-\mu}$, respectively. The element $K_{zx}^\mathrm{int,\:ni}$ can be obtained from $K_{xz}^\mathrm{int,\:ni}$ by swapping $A$ to $B$ and $v_\mu$ to $v^*_{-\mu}$.

The applicability conditions for the answer for the noninvariant part of the interference contribution to the IEI  is similar to Eqs. \eqref{eq: ineq} and \eqref{Tcond: K-i}: $\overline{R} \gg \xi_\mu/\sin^2 \theta_\mu$ and $(\overline{x}_{AB}^2 \sin^2 \theta_\mu + y_{AB}^2)^{1/2} / \xi \ll |M|/(\pi T)$. Typically, the noninvariant part of the interference contribution to the IEI is smaller than the invariant part. However,  if one of the impurities is situated strictly at the edge, such that $R=\overline{R}$ and $|x_{AB}|=|\overline{x}_{AB}|$, the spatial decay of the noninvariant part is exactly the same as the spatial decay of the invariant part. Next, as one can check, in the case of both impurities located exactly at the boundary, $x_A=x_B=0$, the invariant and noninvariant parts of the interference contribution to the IEI compensate each other. Therefore, for the case of both impurities situated exactly at the edge one needs to compute the asymptotic expressions for $K_{ab}^\mathrm{int}$ more accurately.

\subsubsection{Magnetic impurities situated at the edge}

Within the second order expansion in the steepest descent method we calculate the interference contribution to the IEI for two impurities which are located strictly at the edge, $x_A=x_B=0$ (see Appendix \ref{App3}). The corresponding Hamiltonian reads
\begin{widetext}
\begin{align}
\label{eq: eb-IEI}
{H}^{\rm int}_{\rm IEI}=& -\frac{4 \xi \cos\theta_\mu}{y_{AB}} F_\mu(y_{AB}) \Bigl[J_m^AJ_m^B\Bigl( \cos \left(k_F y_{AB}\right)\left(\bm{S}^A_{||}\cdot\bm{S}^B_{||}\right)-\sin \left(k_F y_{AB}\right)\left[\bm{S}^A\times\bm{S}^B\right]_z\Bigr)  \notag \\
& - \Bigl ( 4 J_0^AJ_0^BS^A_xS^B_x+2 \left(J_0^AJ_z^BS_x^AS_z^B+J_0^BJ_z^AS_x^BS_z^A\right)+J_z^AJ_z^BS_z^AS_z^B\Bigr)\cos \left(k_F y_{AB}+2\theta_\mu\right) \Bigr ].
\end{align}
\end{widetext}
We mention that the power-law dependence of the result \eqref{eq: eb-IEI} on the distance is $y_{AB}^{-5/2}$ rather than 
$y_{AB}^{-3/2}$. The additional power is due to the next order expansion in the steepest descent method. At finite temperature the condition of applicability of the result \eqref{eq: eb-IEI} is similar to that for the contribution due to the edge states, $|y_{AB}|/\xi \ll |M|/(\pi T)$.

For impurities situated close to the edge, $|x_A|, |x_B| \ll \xi$, the interference contribution to the IEI is given as a sum of the results \eqref{eq: K-i} and \eqref{eq: K-n-i} as well as the generalization of the result \eqref{eq: eb-IEI}. It has features similar to the result \eqref{eq: eb-IEI}: the exponential decay at the length scale $\xi_\mu$ as well as oscillations with the spatial period $2\pi/k_F$.

\section{Discussion and conclusions\label{Sec:DisConc}}

The results for the IEI reported above was derived  within the lowest order in the exchange coupling constants $J_0$, $J_1$, $J_2$, and $J_m$. The typical value of the IEI is given by the energy scale $T_* \sim  \max \{J^2_z,J^2_0,J^2_m\}/(|M|\xi^4)$ which can be estimated to be of the order of $10^{-3} \div 10^{-4}$~K for the manganese impurities in the CdTe/HgTe/CdTe QW with the width $d=7$ nm \cite{Kurilovichi2016}. 
For the validity of our perturbative calculation the following inequality has to be satisfied, $T_*/|M|\ll 1$. 
In Ref. \cite{Kurilovichi2016} the ratio $T_*/|M|$ was estimated to be of the order of  $10^{-3}$ for the case mentioned above. Such estimate guarantees  validity of the perturbation theory in the exchange interaction.

In the presence of the helical edge states in a 2D TI the IEI between magnetic impurities is determined by the three physically different contributions: contribution due to bulk states (see Eq. \eqref{eq:bulk-IEI}), contribution due to edge states (see Eq. \eqref{Eq: IEI-edge}) and contribution due to interference between bulk and edge states (see Eqs. \eqref{eq: K-i} and \eqref{eq: K-n-i}). With exponential accuracy the spatial dependence of these three contributions can be estimated as   
\begin{equation}
\begin{split}
    H^{\mathrm{bulk}}_{\mathrm{IEI}}\sim e^{-{2R}/{\xi}}, & \quad H^{\mathrm{edge}}_{\mathrm{IEI}}\sim e^{2\overline{x}_{AB}/\xi}, \\
     H^{\mathrm{int}}_{\mathrm{IEI}}& \sim e^{{\overline{x}_{AB}}/{\xi}-{R}/{\xi_\mu}} .
\end{split}
\end{equation}

We note that for the  CdTe/HgTe/CdTe QW with the width $d=7$ nm the decaying length $\xi$ was estimated to be about 20 nm \cite{Kurilovichi2016}. We mention that $\xi$ is much larger than the decaying length for the IEI in a 3D bulk CdTe crystal which is known to be equal to $0.1 \div 1$ nm \cite{Bloembergen1955}. Contrary to $\xi$, the other decaying length, $\xi_\mu$, depends on the chemical potential $\mu$ and $\xi_\mu$ can be much larger than $\xi$ for $|M|-|\mu| \ll |M|$. 

\begin{figure*}[t]
\centerline{\includegraphics[width=0.39\textwidth]{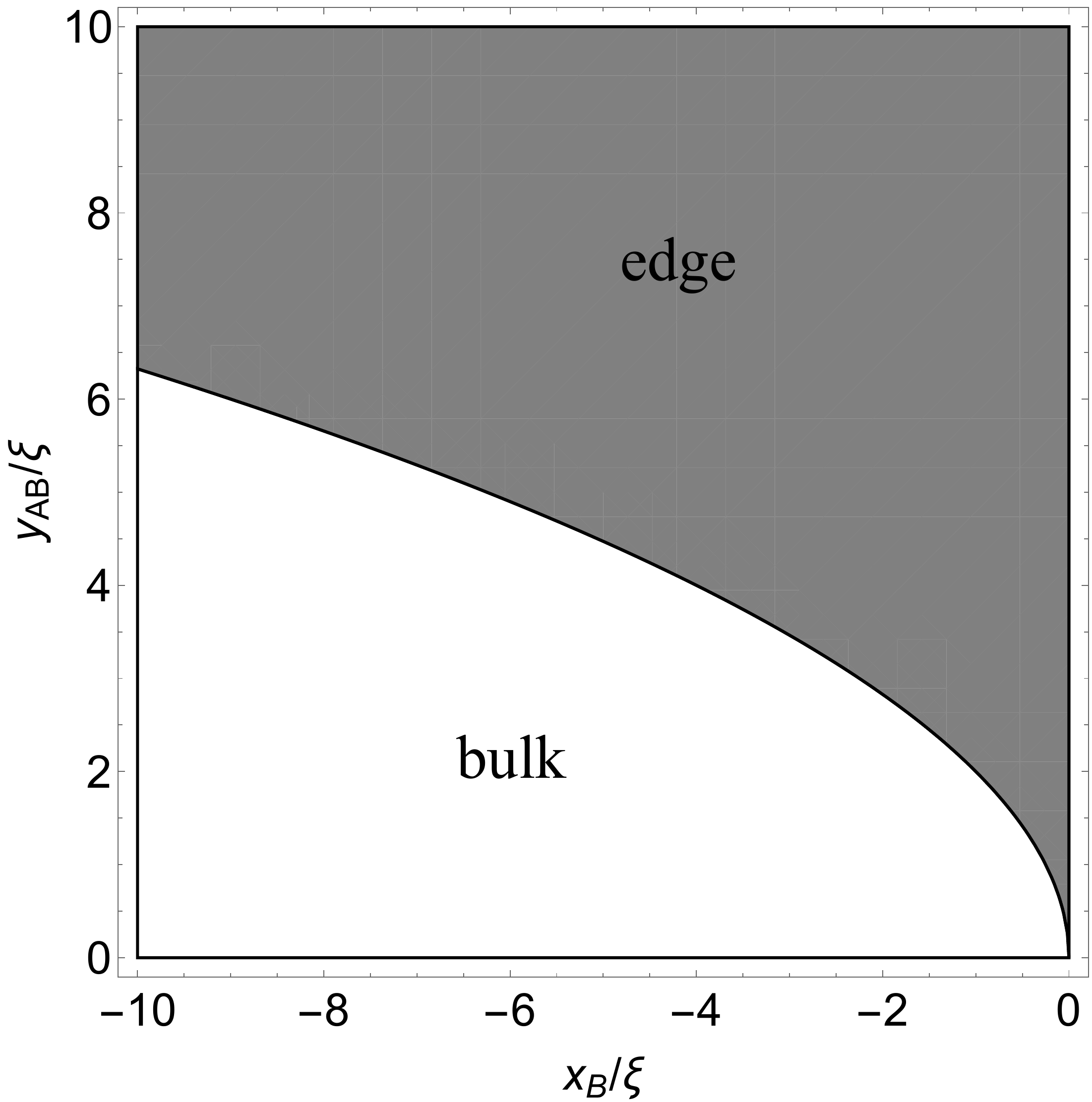}\qquad \includegraphics[width=0.39\textwidth]{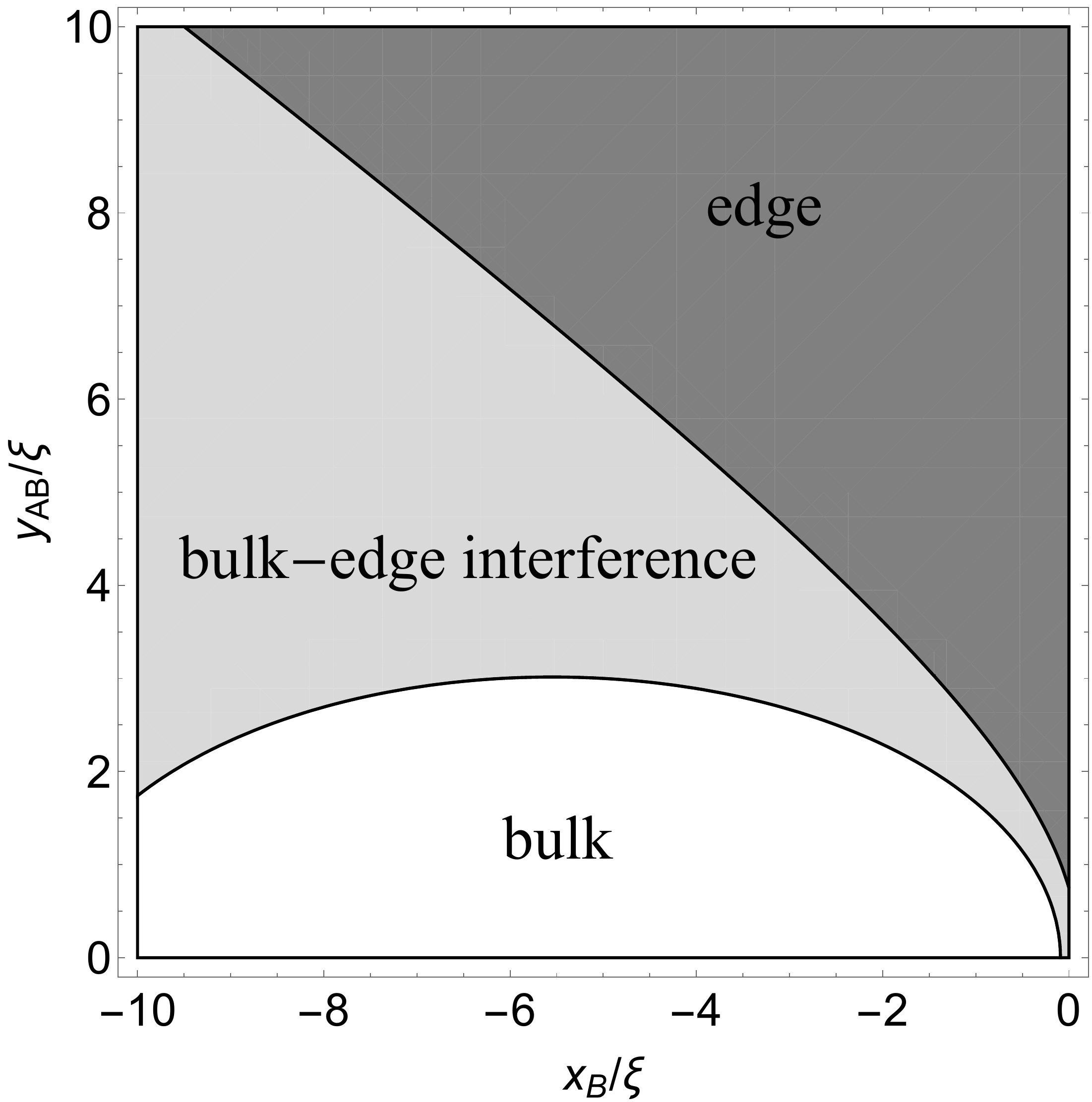}}
\caption{The regions in which different contributions to the IEI are dominant are presented for the case of $\mu = 0$ (the left figure) and $\mu = 0.6 |M|$ (the right figure) for different positions of the impurity B. The impurity A is situated at $|x_A|=\xi$. The white colour depicts the dominance of the IEI meditated by the bulk states, the grey colour indicates the region in which  the interference contribution to the IEI is dominant, and the dark grey colour denotes the dominance of the IEI mediated by the edge states.}
\label{Figure1}
\end{figure*}

In the case of  two impurities situated deep in the bulk, far away from the edge, the bulk contribution to the IEI dominates.  
For impurities which are placed near the edge of a 2D TI the main contribution to the IEI is provided by the edge states. However, this edge contribution to the IEI couples only the in-plane components of the impurity spins \cite{LeeLee2015}. 
At the same time, the interference contribution to the IEI between magnetic impurities situated at the edge involves interaction between $z$ components of the impurity spins (see Eq. \eqref{eq: eb-IEI}). Although, this interaction is exponentially suppressed for the distances along the edge which are larger than $\xi_\mu$ and is of the order of $T_*^{(\mu)} = T_* (1-\mu^2/M^2)^{3/2}$, it can become more important then the contribution due to the edge states, Eq. \eqref{Eq: IEI-edge}, in the case of strong on-site easy axis anisotropy, $H_\mathrm{anis}=-D S_z^2$ with $D>0$. The easy axis anisotropy constricts  spins to be aligned along the $z$ axis with $\bm{S}_{\parallel}=0$. In Ref. \cite{Kurilovichi2016} the on-site anisotropy was estimated to be  $10^3 \div 10^5$ times larger than $T_*$.  Since the IEI between $z$ components of spins is oscillating function of the distance with the period $2\pi/k_F$ we expect formation of a spin-glass state below the temperature $T_*^{(\mu)}$ for randomly distributed magnetic impurities with the 1D density larger than $1/\xi_\mu$. 

Although effects, caused by on-site anisotropy might be crucial, the interference contribution to the IEI can be dominant for specific disposition of the impurities even without the anisotropy. Let us consider the following illustrative example: impurity $A$ is located strictly at the edge while the impurity $B$ is displaced at the distance $|x_B|=X$ away from the boundary towards the bulk of a 2D TI. We will suppose that the distance between the impurities along the edge is equal $y_{AB} = X$. In this situation the three different contributions to the IEI can be estimated as 
\begin{equation}
\begin{split}
    H^{\mathrm{bulk}}_{\mathrm{IEI}} \sim e^{-{2\sqrt{2} X}/{\xi}}, & \quad 
    H^{\mathrm{edge}}_{\mathrm{IEI}} \sim e^{-{2X}/{\xi}} ,\\
    H^{\mathrm{int}}_{\mathrm{IEI}}  \sim & e^{-{X}/{\xi}}e^{-{\sqrt{2} X}/{\xi_\mu}} .
    \end{split}
\end{equation}
Provided $|\mu|>{|M|}/{\sqrt{2}}$, the interference contribution to the IEI has the smallest decaying length and, therefore, dominates over bulk and edge contributions.

To illustrate the importance of the interference term further, we consider the following situation: the impurity $A$ situated in the bulk at some arbitrary fixed distance $x_A$ from the edge whereas the impurity $B$ can be located anywhere. In this situation for $\mu = 0$ the IEI is always dominated either by the bulk or by the edge contribution. Indeed, this follows from estimates:
\begin{align}
    {H^{\mathrm{bulk}}_{\mathrm{IEI}}}/{H^{\mathrm{int}}_{\mathrm{IEI}}}\sim {H^{\mathrm{int}}_{\mathrm{IEI}}}/{H^{\mathrm{edge}}_{\mathrm{IEI}}} .
    \notag
\end{align}

However, for $\mu\ne 0$, the decaying length of the interference contribution to the IEI  increases in comparison with the case of $\mu=0$. For some positions of the impurity $B$, the interference contribution can become the most significant. The comparison of the exponential factors for different positions of the impurity $B$ at a given position of the impurity $A$ is shown in the Fig. \ref{Figure1}. The figure illustrates that for non-zero value of $\mu$ there exists the region for which  the interference contribution to the IEI  is dominant. This area separates the region in which the IEI is mostly due to the bulk states from the region where the interaction due to the edge states is dominant.

Finally, we mention that although the characteristic energy scale $T_*$ of the IEI is rather small, nevertheless, the fine structure of energy levels of a pair of magnetic impurities caused by the IEI can be probed experimentally by broadband electron spin resonance technique coupled with an optical detection scheme \cite{Laplane2016}.

To summarize, we studied the IEI between magnetic impurities near the edge of a 2D topological insulator. This interaction can be divided into three physically different contributions. The first contribution is the IEI mediated by the virtual interband transitions of the bulk electron states. It decays exponentially with the distance between the impurities and has two parts: a rotationally invariant part which was analysed previously in Ref. \cite{Kurilovichi2016} in detail and the part which is not invariant under in-plane rotations. The latter appears if we take into account the change of the bulk states due to the presence of the edge. The second contribution is the RKKY interaction between the impurities due to the helical edge states of a 2D topological insulator. In accordance with the general expectations this contribution decays with distance between the impurities as a power law and oscillates with the period $\pi/k_F$. This contribution  is suppressed if both impurities are situated deep in the bulk. This edge contribution couples only in-plane components of the impurity spins. Finally, the last contribution to the IEI can be interpreted as the interaction, mediated by the interference between the bulk and  edge states. This term oscillates with  $k_F$ and decays exponentially with the distance between the impurities. Interestingly, the decaying length of this interference contribution is controlled by the position of the chemical potential within the bulk gap. This fact makes the interference contribution to the IEI to be dominant in the case of some  specific disposition of magnetic impurities.

\begin{acknowledgments}

We thank B. Aronson, M. Durnev, M. Feigel'man, Y. Gefen, M. Glazov,  M. Goldstein, G. Min'kov, I. Rozhansky and, especially, S. Tarasenko, for useful discussions. The work was partially supported by the Russian Foundation for Basic Research under the Grant No. 15-52-06005, Russian President Grant No. MD-5620.2016.2, and Russian President Scientific Schools Grant NSh-10129.2016.2.

\end{acknowledgments}

\appendix

\section{The contribution to the IEI due to bulk states\label{App1}}

In this appendix we present details of the calculation of the contribution due to bulk states to the IEI. Using Eqs. \eqref{H-IEI}, we find the following expression valid at zero temperature
\begin{align}
H_{\mathrm{IEI}}^{\mathrm{bulk}} & = \frac{1}{4}\sum_{s,s'=\pm}\int\frac{d\varepsilon}{2\pi}\frac{d k_x}{2\pi}\frac{d k_y}{2\pi}\frac{d z_x}{2\pi}\frac{d z_y}{2\pi}e^{i(k_y-z_y)y_{AB}}\notag
\\
& \times \frac{\Tr [\mathcal{J}^A \mathcal{B}_{s}(\bm{k},x_A,x_B) \mathcal{J}^B \mathcal{B}_{s'}(\bm{z},x_B,x_A)]}{[i\varepsilon+\mu-s\mathcal{E}( k)][i\varepsilon+\mu-s'\mathcal{E}( z)]} .
\end{align}
Assuming that the chemical potential is pinned within the bulk gap, we can integrate over energy $\varepsilon$ and obtain
\begin{align}
H_{\mathrm{IEI}}^{\mathrm{bulk}} = - & \frac{1}{4}\sum_{s'=\pm} \int \limits_0^\infty dt  
\Tr [\mathcal{J}^A \tilde{\mathcal{B}}_{s}(x_A,x_B,y_{AB})\notag \\
& \times  \mathcal{J}^B \tilde{\mathcal{B}}_{-s}(x_B,x_A,y_{BA})] .
\end{align}
Here we introduced integration over an auxiliary variable $t$ and 
\begin{gather}
\tilde{\mathcal{B}}_{s}(x_A,x_B,y_{AB}) = \int \frac{d^2\bm{k}}{(2\pi)^2} e^{i k_y y_{AB} - t \mathcal{E}(k)} \mathcal{B}_{s}(\bm{k},x_A,x_B) .
\label{eq:tildeB1}
\end{gather}
To proceed further, we need to evaluate integral over momentum $\bm{k}$. Since we are interested in the asymptotic behaviour of the IEI at large distance, it is enough to evaluate the integral over momentum in the saddle-point approximation. In particular, we shall use the following general result
\begin{equation}
\int\frac{d^2\bm{q}}{(2\pi)^2}F(\bm q) e^{i\bm q \bm r - u \sqrt{1+q^2}}\approx\frac{u F(\bm{q}_0)}{2\pi (r^2+u^2)}e^{-\sqrt{r^2+ u^2}},
\label{eq:SPint}
\end{equation}
where $\bm{q}_0 = i \bm{r}/\sqrt{r^2+u^2}$. This result is valid provided $\sqrt{r^2+u^2}\gg 1$. With the help of Eq. \eqref{eq:SPint}, one finds
\begin{gather}
\tilde{\mathcal{B}}_{s}(x_A,x_B,y_{AB}) = s \begin{pmatrix}
\tilde{I}_{1,s} & \tilde{I}_{2,s} & 0 & 0\\
-\tilde{I}_{2,-s}^*& -\tilde{I}_{1,-s}^*& 0 & 0\\
0 & 0 & \tilde{I}_{1,s}^*  & \tilde{I}_{2,s}^*\\
0 & 0 &  -\tilde{I}_{2,-s} & -\tilde{I}_{1,-s}\\
\end{pmatrix} .
\label{eq:exp:B+}
\end{gather}
Using the relation $\tilde{\mathcal{B}}_{s}(x_B,x_A,y_{BA}) =\tilde{\mathcal{B}}^\dag_{s}(x_A,x_B,y_{AB})$, one can obtain the expression for $\tilde{\mathcal{B}}_{-s}(x_B,x_A,y_{BA})$ from Eq. \eqref{eq:exp:B+} by substituting $-\tilde{I}_{1,-s}^*$ for $\tilde{I}_{1,s}$ and vice versa. 
Here, the functions $\tilde{I}_1$ and $\tilde{I}_2$ are given as follows 
\begin{align}
\begin{pmatrix}
\tilde{I}_{1,s} \\
\tilde{I}_{2,s} 
\end{pmatrix}
& =
\begin{pmatrix}
 - 1 + \frac{A t s}{\sqrt{R^2+A^2 t^2}} \\
\frac{i n_+  R}{\sqrt{R^2+A^2 t^2}} 
\end{pmatrix} \frac{e^{-\sqrt{R^2+A^2 t^2}/\xi} }{2\pi \xi (R^2+A^2 t^2)^{1/2}} 
\notag \\
& +
\begin{pmatrix}
i\Bigl (\frac{\overline{R}\nu_+}{\sqrt{\overline{R}^2+A^2t^2}}+ \frac{s At}{\sqrt{\overline{R}^2+A^2t^2}}-1\Bigr )^2 \\
\Bigl |\frac{\overline{R}\nu_+}{\sqrt{\overline{R}^2+A^2t^2}} -1 \Bigr |^2 - \frac{2 i \nu_y s At \overline{R}}{\overline{R}^2+A^2t^2} - \frac{A^2t^2}{\overline{R}^2+A^2t^2}
\end{pmatrix}
\notag \\
& \times   \frac{e^{-\sqrt{\overline{R}^2+A^2t^2}/\xi}}{4\pi \xi (\nu_y \overline{R}-i s A t)}  .
\label{eq:I1I2}
\end{align}
We note that under the interchange of the points $\bm{R}_A$ and $\bm{R}_B$ the functions $\tilde{I}_{1,s}$ and $\tilde{I}_{2,s}$ transfer to $\tilde{I}_{1,s}^*$ and $-\tilde{I}_{2,s}$, respectively. To perform integration over $t$, we can simplify expressions for $\tilde{I}_{1,s}$ and $\tilde{I}_{2,s}$ by
expanding in $t$ the square root in the exponents and to neglect $t$  in all other places:
\begin{align}
\begin{pmatrix}
\tilde{I}_{1,s} \\
\tilde{I}_{2,s} 
\end{pmatrix} & \approx \begin{pmatrix}
I_1 \\
I_2 
\end{pmatrix}
 = \begin{pmatrix}
-1 \\
i n_+ 
\end{pmatrix} \frac{1}{2\pi \xi R} e^{-R/\xi - A|M| t^2/2R}
\notag \\
 +&
\begin{pmatrix}
i (\nu_+-1)^2 \\
|\nu_+-1|^2
\end{pmatrix} \frac{1}{4\pi \xi \nu_y \overline{R}} \,  e^{-\overline{R}/\xi - A |M|t^2/2\overline{R}} 
\label{eq:I1I2-s}.
\end{align}
This is allowed provided the following inequalities hold
\begin{equation}
\frac{R}{\xi} \gg t |M|  \gg1, \qquad \frac{\overline{R}}{\xi} \gg t |M|  \gg 1 .
\end{equation}
Then using Eqs. \eqref{eq:exp:B+} and \eqref{eq:I1I2-s}, we integrate over $t$ (notice, that scale of convergence of the corresponding integrals over $t$ makes inequalities above well justified provided $R/\xi \gg 1$, $\bar{R}/\xi \gg 1$) and obtain  
\begin{equation}
H_{\mathrm{IEI}}^\mathrm{bulk}=\sum_{a,b=x,y,z}S^A_a K^{\mathrm{bulk}}_{ab}S^B_b ,
\end{equation}
where
\begin{align}
&K^{\mathrm{bulk}}_{xx}=J_m^AJ_m^B \mathcal{F}_2-2J_0^AJ_0^B (\mathcal{F}_2+\mathcal{F}_3)+\mathrm{c.c.} ,\notag\\
&K^{\mathrm{bulk}}_{xy}=-iJ_m^AJ_m^B \mathcal{F}_2-2iJ_0^AJ_0^B (\mathcal{F}_2-\mathcal{F}_3)+\mathrm{c.c.} ,\notag\\
&K^{\mathrm{bulk}}_{xz}=-2iJ_0^AJ_z^B\mathcal{F}_6+\mathrm{c.c.} ,\notag\\
&K^{\mathrm{bulk}}_{yx}=iJ_m^AJ_m^B \mathcal{F}_2+2iJ_0^AJ_0^B (\mathcal{F}_2+\mathcal{F}_3)+\mathrm{c.c.} ,\notag\\
&K^{\mathrm{bulk}}_{yy}=J_m^AJ_m^B \mathcal{F}_2-2J_0^AJ_0^B (\mathcal{F}_2-\mathcal{F}_3)+\mathrm{c.c.} ,\notag\\
&K^{\mathrm{bulk}}_{yz}=2J_0^AJ_z^B\mathcal{F}_6+\mathrm{c.c.} ,\notag\\
&K^{\mathrm{bulk}}_{zx}=2iJ_z^AJ_0^B\mathcal{F}_5+\mathrm{c.c.} ,\notag\\
&K^{\mathrm{bulk}}_{zy}=2J_z^AJ_0^B\mathcal{F}_5+\mathrm{c.c.} ,\notag\\
&K^{\mathrm{bulk}}_{zz}=2\left( J_1^A(\mathcal{F}_1 J_1^B+\mathcal{F}_4 J_2^B)+J_2^A(\mathcal{F}_4 J_1^B+\mathcal{F}_1J_2^B)\right) .\label{eq:Kbulk}
\end{align}
Here the functions $\mathcal{F}_{1,\dots, 6}$ are defined as follows:
\begin{widetext}
\begin{align}
  &\mathcal{F}_1= \frac{1}{2} \int\limits_0^\infty dt |I_1|^2 = \mathcal{F}(R,R)+\frac{\overline{R}^2}{4 y_{AB}^2}|\nu_+ - 1|^4\mathcal{F}(\overline{R},\overline{R})+i\frac{\overline{R}}{2y_{AB}}\left((\nu_- - 1)^2-(\nu_+-1)^2\right)\mathcal{F}(R,\overline{R}) , \notag\\
  &\mathcal{F}_2= \frac{1}{2} \int\limits_0^\infty dt I_1^2 =\mathcal{F}(R,R)-\frac{\overline{R}^2}{4 y_{AB}^2}\left(\nu_+ - 1\right)^4\mathcal{F}(\overline{R},\overline{R})-i\frac{\overline{R}}{y_{AB}}(\nu_+-1)^2\mathcal{F}(R,\overline{R}) , \notag\\
  &\mathcal{F}_3 =- \frac{1}{2} \int\limits_0^\infty dt I_2^2 =n_+^2\mathcal{F}(R,R)-\frac{\overline{R}^2}{4 y_{AB}^2}|\nu_+ - 1|^4\mathcal{F}(\overline{R},\overline{R})-i\frac{\overline{R}}{y_{AB}}n_+|\nu_+-1|^2\mathcal{F}(R,\overline{R}) , \notag\\
  &\mathcal{F}_4=\frac{1}{2} \int\limits_0^\infty dt |I_2|^2 = \mathcal{F}(R,R)+\frac{\overline{R}^2}{4 y_{AB}^2}|\nu_+ - 1|^4\mathcal{F}(\overline{R},\overline{R}))+i\frac{\overline{R}}{2y_{AB}}(n_+-n_-)|\nu_+-1|^2\mathcal{F}(R,\overline{R}) ,
  \notag
   \end{align}
  \begin{align}
  &\mathcal{F}_5=\frac{1}{2} \int\limits_0^\infty dt I_1^* I_2 = -i n_+\mathcal{F}(R,R)-i\frac{\overline{R}^2}{4 y_{AB}^2}\left(\nu_- - 1\right)^2|\nu_+ - 1|^2\mathcal{F}(\overline{R},\overline{R}))+\frac{\overline{R}}{2y_{AB}}\left(n_+(\nu_- - 1)^2-|\nu_+-1|^2\right)\mathcal{F}(R,\overline{R}) , \notag \\
   &\mathcal{F}_6=- \frac{1}{2} \int\limits_0^\infty dt I_1^* I_2^* = -i n_-\mathcal{F}(R,R)+i\frac{\overline{R}^2}{4 y_{AB}^2}\left(\nu_- - 1\right)^2|\nu_+ - 1|^2\mathcal{F}(\overline{R},\overline{R}))+\frac{\overline{R}}{2y_{AB}}\left(n_-(\nu_- - 1)^2+|\nu_+-1|^2\right)\mathcal{F}(R,\overline{R}) . 
 \end{align}
\end{widetext}
The function $\mathcal{F}(R,R^\prime)$ describes the exponential decay of the IEI
\begin{equation}
\mathcal{F}(R,R^\prime)=\frac{|M|^3}{16 A^4}\sqrt{\frac{2\xi^3}{\pi^3 R R^\prime (R+R^\prime)}} \, e^{-R/\xi-R^\prime/\xi} .
\end{equation}
The result \eqref{eq:Kbulk} is valid provided the following inequalities are fulfilled:
\begin{equation}
R \gg \xi, \qquad \overline{R} \gg \xi .
\end{equation}
In the case of both impurities located in the bulk far away from the boundary, $|x_A|, |x_B| \gg \xi$, i.e. the distance 
$R\ll \overline{R}$, then the result \eqref{eq:Kbulk} transforms into the expression \eqref{eq:bulk-IEI}.

\section{Interference contribution to the IEI \label{App2}}

In this Appendix we present details  of   derivation of the interference contribution to the IEI. Using Eqs. \eqref{eq:G:edge:2} and \eqref{eq: bulkgf},  we can express the interference contribution to the IEI at zero temperature and for $y_{AB}>0$ as follows
\begin{gather}
H_{\mathrm{IEI}}^{\mathrm{int}} = \frac{i |M|}{2 A^2} e^{{\overline{x}_{AB}}/{\xi}}\sum_{s,s^\prime=\pm} \int  \frac{d\varepsilon}{2\pi} \int \frac{d^2\bm{k}}{(2\pi)^2} \frac{ s^\prime \theta(-s^\prime \varepsilon)}{i\varepsilon+\mu - s\mathcal{E}(k)} \notag \\
\times e^{{s^\prime (\varepsilon- i \mu) y_{AB}}/{A}}\Bigl [ e^{i k_y y_{AB}} \Tr  \mathcal{J}^A \mathcal{B}_s(\bm{k},x_A,x_B) \mathcal{J}^B \Gamma_{-s^\prime} \notag \\+ e^{i k_y y_{BA}} \Tr \mathcal{J}^A \Gamma_{s^\prime}  \mathcal{J}^B \mathcal{B}_s(\bm{k},x_B,x_A) \Bigr ] .
\label{eq:app:int:3}
\end{gather}
After introducing an integration over a variable $t$ to raise the denominator $i\varepsilon+\mu - s\mathcal{E}(k)$  into the exponent, we can integrate over $\varepsilon$ and obtain
\begin{gather}
H_{\mathrm{IEI}}^{\mathrm{int}} = -\frac{|M|}{4\pi A^2} e^{{\overline{x}_{AB}}/{\xi}}\sum_{s,s^\prime=\pm} \int  \limits_0^\infty dt  \, 
\frac{e^{i s^\prime k_F y_{AB}+ s \mu t}}{t + i ss^\prime {y_{AB}}/{A}} \notag \\
\times  \Tr \Bigl [ \mathcal{J}^A \tilde{\mathcal{B}}_s(x_A,x_B,y_{AB}) \mathcal{J}^B \Gamma_{s^\prime}\notag \\
 +\mathcal{J}^A \Gamma_{-s^\prime}  \mathcal{J}^B \tilde{\mathcal{B}}_s(x_B,x_A,y_{BA})  \Bigr ] .
\end{gather}
Next, we find
\begin{gather}
H_{\mathrm{IEI}}^{\mathrm{int}} = \sum_{s^\prime=\pm} \Tr \Bigl [ \mathcal{J}^A \hat{\mathcal{B}}_{s^\prime}(x_A,x_B,y_{AB}) \mathcal{J}^B \Gamma_{s^\prime} \notag \\
+\mathcal{J}^B \hat{\mathcal{B}}^\dag_{s^\prime}(x_A,x_B,y_{AB}) \mathcal{J}^A \Gamma_{s^\prime} \Bigr ] ,
\end{gather}
where
\begin{gather}
\hat{\mathcal{B}}_{s^\prime}(x_A,x_B,y_{AB})  = -\frac{|M|}{4\pi A^2} e^{{\overline{x}_{AB}}/{\xi}}\sum_{s=\pm} \int  \limits_0^\infty dt  \notag \\
\times
\frac{e^{i s^\prime k_F y_{AB}+ s \mu t}}{t + i ss^\prime {y_{AB}}/{A}}\, \tilde{\mathcal{B}}_s(x_A,x_B,y_{AB}) .
\label{eq:I1I2-1}
\end{gather}
After inspection of Eq. \eqref{eq:I1I2}, we see that one can evaluate the integral over $t$ within the saddle point approximation, provided $R$ and  $\overline{R}$ are large enough. We note that in the sum over $s$ the term with $s= \sgn \mu$ yields the leading contribution only. In particular, we use the following result for $1>a>0$
\begin{equation}
\int \limits_0^\infty du F(u) e^{a u - \sqrt{r^2+u^2}} \approx \frac{\sqrt{2\pi r}}{(1-a^2)^{3/4}} F(u_0) e^{-r \sqrt{1-a^2}} , 
\label{eq:app:int:4}
\end{equation}
where $u_0 = a r/\sqrt{1-a^2}$. This saddle-point result is valid provided $r \gg 1/(a^2\sqrt{1-a^2})$. In terms of $R$ and $\overline{R}$ this condition implies that $R, \overline{R}  \gg \xi (M^3/\mu^2 \sqrt{M^2-\mu^2})$.

Performing integration over $t$ in Eq. \eqref{eq:I1I2-1} with the help of the saddle point result \eqref{eq:app:int:4}, we find
\begin{gather}
 \hat{\mathcal{B}}_{s^\prime}(x_A,x_B,y_{AB}) = 
\frac{e^{i s^\prime k_F y_{AB}}}{\sin\theta_\mu + i s^\prime n_y \cos\theta_\mu} F_\mu(R) \notag \\
\times 
 \begin{pmatrix}
1-\sin\theta_\mu& -i n_+ \cos \theta_\mu & 0 & 0 \\
 -i n_- \cos \theta_\mu & -1-\sin \theta_\mu & 0 & 0 \\
 0 & 0& 1-\sin\theta_\mu & i n_- \cos \theta_\mu \\
 0 & 0 & i n_+ \cos \theta_\mu & -1-\sin \theta_\mu
 \end{pmatrix} 
 \notag \\
 -\frac{e^{i s^\prime k_F y_{AB}} F_\mu(\overline{R}) }{\sin\theta_\mu + i s^\prime \nu_y \cos\theta_\mu} 
 \begin{pmatrix}
u_\mu& v_\mu & 0 & 0 \\
 -v^*_{-\mu} & -u^*_{-\mu} & 0 & 0 \\
 0 & 0& u^*_\mu & v^*_\mu \\
 0 & 0 & -v_{-\mu} & -u_{-\mu}
 \end{pmatrix} ,
 \label{eq: B-hat}
 \end{gather}
Next, performing summation over $s^\prime$, we obtain the results \eqref{eq: K-i} and \eqref{eq: K-n-i}: the first term in \eqref{eq: B-hat} results in the invariant part of the interference IEI \eqref{eq: K-i}, while the second term - in the noninvariant part \eqref{eq: K-n-i} (see \eqref{eq: sum-h}).

\section{Interference contribution to the IEI between the impurities situated exactly at the edge \label{App3}}

In this Appendix we present details of derivation of the interference contribution to the IEI for magnetic impurities situated exactly at the edge of the 2D topological insulator, i.e. for the case $x_A=x_B=0$. In this case the expressions \eqref{eq:I1I2} for the integrals $
\tilde{I}_{1,s}$ and $\tilde{I}_{2,s}$ vanish identically. Therefore, one has to compute the integrals over $\bm{k}$ more accurately. We find that 
\begin{equation}
\tilde{I}_{1,s} = i \tilde{I}_{2,s} = \int \frac{d^2\bm k}{(2\pi)^2} e^{i s k_y y_{AB} - t \mathcal{E}(k)} \frac{s A^2 k_x^2}{\mathcal{E}(k) [\mathcal{E}(k) + A k_y ]} .
\end{equation}
Evaluating integral over $\bm{k}$ in the saddle-point approximation, we find
\begin{gather}
\tilde{I}_{1,s} = i \tilde{I}_{2,s} = \frac{s}{2\pi} \frac{1}{A t + i s y_{AB}}\,\frac{e^{-\sqrt{y_{AB}^2+A^2 t^2}}}{\sqrt{y_{AB}^2+A^2 t^2}} .
\end{gather}
Performing integration over $t$ in Eq. \eqref{eq:I1I2-1} with the help of Eq. \eqref{eq:app:int:4}, we find
\begin{gather}
 \hat{\mathcal{B}}_{s^\prime}(0,0,y_{AB}) = 
s^\prime e^{i s^\prime k_F y_{AB}+i s^\prime \theta_\mu} \frac{\xi \cos\theta_\mu}{ y_{AB}} F_\mu(y_{AB}) \notag \\
\times 
\begin{pmatrix}
e^{i \theta_\mu} & - i e^{i \theta_\mu} & 0 & 0\\
i e^{i \theta_\mu} & e^{i \theta_\mu} & 0 & 0 \\
0 & 0 & - e^{-i \theta_\mu} & - i e^{-i \theta_\mu} \\
0 & 0& i e^{-i \theta_\mu}  &- e^{-i \theta_\mu} 
\end{pmatrix} .
\end{gather}

Using this expression, we find the result \eqref{eq: eb-IEI} for the interference contribution to the IEI for the case of magnetic impurities situated exactly at the edge.


\bibliography{biblio}

\end{document}